# Dielectric response of soft mode in ferroelectric SrTiO$_3$

**Jiaguang Han**
*School of Electrical and Computer Engineering, Oklahoma State University, Stillwater, Oklahoma 74078, and Shanghai Institute of Applied Physics, Chinese Academy of Sciences, Shanghai 201800, People's Republic of China*

**Fan Wan and Zhiyuan Zhu**
*Shanghai Institute of Applied Physics, Chinese Academy of Sciences, Shanghai 201800, People's Republic of China*

**Weili Zhang[a]**
*School of Electrical and Computer Engineering, Oklahoma State University, Stillwater, Oklahoma 74078*

We report far-infrared dielectric properties of powder form ferroelectric SrTiO$_3$. Terahertz time-domain spectroscopy (THz-TDS) measurement reveals that the low-frequency dielectric response of SrTiO$_3$ is a consequence of the lowest transverse optical (TO) soft mode TO1 at 2.70 THz (90.0 cm$^{-1}$), which is directly verified by Raman spectroscopy. This result provides a better understanding of the relation of low-frequency dielectric function with the optical phonon soft mode for ferroelectric materials. Combining THz-TDS with Raman spectra, the overall low-frequency optical phonon response of SrTiO$_3$ is presented in an extended spectral range from 6.7 cm$^{-1}$ to 1000.0 cm$^{-1}$.

[a] Electronic mail: wwzhang@okstate.edu

Since the discovery of the polar soft mode in Strontium Titanate ($SrTiO_3$) crystals, there has been a great interest and challenge to understand the behaviors of dielectric properties of ferroelectric materials, particularly to explore the relationship between the dielectric properties and the optical phonon responses.[1-3] As an incipient ferroelectric material $SrTiO_3$ of the perovskite structure plays a very important role among the ferroelectrics family, $ABO_3$ compounds. Because of the remarkably high dielectric permittivity, tunability and low microwave loss, $SrTiO_3$ becomes very attractive and popular not only for fundamental research, but also for device applications. It is an important material in high-storage-density capacitor structures, such as dynamic random access memory because of high static dielectric constant.[4] $SrTiO_3$ has also been utilized in various tunable microwave devices.[5,6] In addition, the unique characteristics of $SrTiO_3$ make it a material of choice for applications in THz devices, particularly for controlling THz radiation and devices.[7] The dielectric properties of $SrTiO_3$ are closely related to its low-frequency optical phonon modes.

The recent advance in highly sensitive Terahertz time-domain spectroscopy (THz-TDS) has provided us with a powerful modality to explore low-frequency properties of materials, especially for the investigation of low-frequency optical phonon modes.[8-12] In this letter, the complex dielectric function along with power absorption and refractive index of powder form $SrTiO_3$ sample are measured in the frequency range from 0.2 THz (6.7 cm$^{-1}$) to 2.0 THz (66.7 cm$^{-1}$) by use of THz-TDS. The experimental results are well fit by the classical pseudo-harmonic model. To verify the THz-TDS measurements, the optical phonon resonance of $SrTiO_3$ is also characterized by Raman light scattering. Our results reveal that the dielectric response of $SrTiO_3$ in the THz region is dominated by the lowest transverse optical (TO) soft mode TO1 at frequency ($\omega_{TO}/2\pi$) 2.70 THz (90.0 cm$^{-1}$). The complex low-frequency optical phonon response processes of SrTiO$_3$ are depicted in a broad spectral range (6.7 – 1000.0 cm$^{-1}$) by combining THz-TDS with Raman spectroscopy. The THz-TDS results of powder form SrTiO$_3$ are compared with those of different forms of SrTiO$_3$, demonstrating that the powder form SrTiO$_3$ shows similar dielectric response behaviors of single-crystalline and thin-film SrTiO$_3$.

The SrTiO$_3$ sample (Sigma-Aldrich) has a purity of higher than 99%, a room-temperature density of 4.81 g/ml, and an average grain size of less than 5 μm. The sample for analysis was prepared by milling the powder carefully and then placed in a cell with two parallel windows made from 636-μm-thick, p-type 20-Ω cm silicon slabs.[11] An identical empty cell was used as a reference. Both cells were mounted on an aluminum holder and centered over a 5-mm-diameter hole to define the optical aperture. The photoconductive switch-based THz-TDS system employed here has a useful bandwidth of 0.1 to 4.5 THz and a signal-to-noise ratio (S/N) > 10,000:1 as described previously.[11,12]

Figure 1 shows the THz-TDS transmission spectra of powder absorption $\alpha(\omega)$ and refractive index $n(\omega)$ of the SrTiO$_3$ sample. In order to increase the S/N, the experimental data are an average of seven individual measurements. The power absorption is enhanced with increasing frequency; no prominent absorption peaks are observed below 2.0 THz as confirmed by the refractive index that shows no prominent changes. The frequency-dependent complex dielectric function of SrTiO$_3$ is determined by the recorded data of the power absorption and the refractive index through the relationship: $\varepsilon(\omega) = (n_r + in_i)^2$, where the imaginary part of the refractive index $n_i$ is related to the power absorption as $n_i = \alpha\lambda/4\pi$. Hence the real and imaginary dielectric constants are determined as $\varepsilon_r = n_r^2 - (\alpha\lambda_0/4\pi)^2$ and $\varepsilon_i = \alpha n_r \lambda_0/2\pi$, respectively.[12] The observed real and imaginary parts of complex dielectric function are shown by open circles in Fig. 2.

Since the measured sample is a composite medium of SrTiO$_3$ and air, the dielectric constant shown in Fig. 2 is an effective quantity $\varepsilon_{eff}$. Hence, the dispersion of the dielectric constant in the THz region is analyzed by use of the Bruggeman theory, one of the most commonly used effective medium approximation (EMA). The Bruggeman EMA is given by[13]

$$f\left(\frac{\varepsilon_m - \varepsilon_{eff}}{\varepsilon_m + 2\varepsilon_{eff}}\right) + (1-f)\left(\frac{\varepsilon_h - \varepsilon_{eff}}{\varepsilon_h + 2\varepsilon_{eff}}\right) = 0 , \qquad (1)$$

where $f$ is the filling factor that defines the volume fraction of SrTiO$_3$. $\varepsilon_h$ and $\varepsilon_m$ are the dielectric constants of the host medium and the pure SrTiO$_3$, respectively. In this case, the host medium is air, giving $\varepsilon_h = \varepsilon_{air} = 1.0$. The dielectric function $\varepsilon_m$ for pure SrTiO$_3$ can be calculated by the simple classical pseudo-harmonic model, where the interaction of a radiation field with the fundamental lattice vibration results in absorption of electromagnetic wave due to the creation or annihilation of lattice vibration. In the frame of the classical pseudo-harmonic model, the complex dielectric function can be expressed as[14]

$$\varepsilon_m(\omega) = \varepsilon_\infty + \frac{(\varepsilon_0 - \varepsilon_\infty)\omega_{TO}^2}{\omega_{TO}^2 - \omega^2 - i\gamma\omega} = \varepsilon_{mr} + i\varepsilon_{mi} , \qquad (2)$$

where $\varepsilon_\infty$ is high frequency dielectric constant, $\varepsilon_0$ is low frequency dielectric constant, $\omega_{TO}$ is the angular frequency of the transverse optical phonon mode, and $\gamma$ is the damping constant.

The solid curves in Figs. 1 and 2 represent the theoretical fitting to the measured power absorption, refractive index and complex dielectric function with the lowest transverse optical soft phonon mode centered at $\omega_{TO}/2\pi = 2.70 \pm 0.1$ THz and with a linewidth $\gamma/2\pi = 1.30 \pm 0.1$ THz. The optical constants used in the fitting are $\varepsilon_\infty = 5.20$ and $\varepsilon_0 = 310$,[2] and the filling factor $f$ is estimated to be 0.31. The parameters are determined with good accuracy by the fitting on the real and imaginary parts of the dielectric function simultaneously. The determined eigenfrequency of the soft mode 2.70 THz is quite consistent with that observed in the bulk crystals and thin films.[2,4,5,15] The good agreement between the experimental data and the theoretical fitting shown in Figs. 1 and 2 indicates that the low-frequency THz response of $SrTiO_3$ is mainly attributed to the lowest soft phonon mode TO1 located at 2.70 THz.

Generally, the phonons of $SrTiO_3$ are triply degenerate $F_{1u}$ modes and the dielectric response is mainly ascribed to the contribution of three ferroelectric transverse optical (TO) modes. The lowest ferroelectric optical soft mode TO1 is induced predominantly by Ti-O-Ti bending; TO2 is mainly due to Sr against TiO6 octahedra translations; TO4 includes Ti-O stretching; and TO3 ($F_{2u}$) is a silent mode and optical inactive.[3,16,17] Both TO2 and TO4 are temperature independent and have a very slightly variation in the magnitude of dielectric strengths and eigenfrequencies when temperature varies. According to the previous experimental reports, the typical eigenfrequencies of TO2 and TO4 modes locate at 178.0 $cm^{-1}$ and 546.0 $cm^{-1}$, respectively. The eigenfrequency of soft mode TO1 depends mostly on temperature and both its strength and eigenfrequency have distinct changes with temperature. Its frequency decreases drastically at lower temperatures, while the magnitude of dielectric constant increases at the mean time. In fact, TO1 is both Raman and IR active; however, it is difficult for conventional IR reflection and Raman techniques to resolve the lowest soft mode, especially at low temperatures. This is because the frequency value of TO1 mode is beyond the reliable measurement range. The previous fit to the infrared and Raman measurements showed that the eigenfrequency of optical soft mode TO1 was near 90.0 $cm^{-1}$ at room temperature,[3,4] which agrees well with our extrapolated results based on the fitting to the THz-TDS data. A further obviously understanding of such phonon features requires direct measurement of the soft TO1 mode. However, THz-TDS does not cover up to 90.0 cm$^{-1}$ due to the fact that the SrTiO$_3$ sample is highly absorptive at frequencies above 2.0 THz (66.7 cm$^{-1}$). Therefore, Raman light scattering is utilized as a complement of THz-TDS to extend the spectral range to higher frequencies.

The Raman spectrum was excited with the 514.5 nm line of an Argon ion laser with a beam spot focused down to 5 μm in diameter. Figure 3 shows a completed optical response processes over a frequency range from 6.7 to 1000.0 cm$^{-1}$ by combining THz-TDS and Raman spectroscopy. The solid curve illustrates the Raman spectrum of the SrTiO$_3$ beginning from 67.0 cm$^{-1}$, while the open circles represent the recorded data by THz-TDS at lower frequencies. The Raman spectrum reveals four broad optical phonon bands that correspond to the active vibrational modes of $F_{1u}$ symmetry: the lowest optical soft mode TO1 at 90.0 cm$^{-1}$, the TO2-LO1 band near 170.0 cm$^{-1}$, the TO4 mode at 539.0 cm$^{-1}$, and the LO4-A2g band at 793.0 cm$^{-1}$. The other abundantly observed features are related to the additional active modes. These optical responses show good consistency with the previous work on SrTiO$_3$ thin film and ceramic samples.[3,7,18,19] The inset of Fig. 3 shows the magnified features of the TO1 mode at frequencies between 80.0 to 130.0 cm$^{-1}$. The Raman spectrum has confirmed that the dielectric properties of SrTiO$_3$ characterized by THz-TDS is indeed dominated by the soft TO1 mode at 90.0 cm$^{-1}$.

Additionally, the THz-TDS results of powder SrTiO$_3$ are compared with that of single-crystalline and thin-film SrTiO$_3$. The <100> oriented single-crystalline SrTiO$_3$ is undoped, with dimensions of 10 × 10 × 0.5 mm$^3$ and high dielectric constant ~300 (MTI Corporation). Figure 4 illustrates a comparison of power absorption and refractive index between different forms of SrTiO$_3$. The measured results of single-crystalline SrTiO$_3$ are well fit by the classical pseudo-harmonic model in Eq. (2), while the data of pure powder are extracted through EMA of Eq.(1) and are fit with Eq. (2). The theoretical fitting indicates that the THz response of single-crystalline SrTiO$_3$ is also dominated by the soft mode TO1 at 90.0 cm$^{-1}$. The data of thin film for comparison are obtained from Ref. [7]. Compared to single-crystalline and thin-film SrTiO$_3$, the powder form shows slightly higher absorption, but all these samples have similar frequency-dependent behaviors. At 0.5 THz, the refractive index of thin film (13.9) and pure powder (18.0) are close to the value 19.2 of single-crystalline SrTiO$_3$, but much higher than 3.0 for the composites (Fig. 1b) due to inclusion of air in the later. The above comparison has shown that the powder form SrTiO$_3$ has similar dielectric properties and low-frequency optical response with that of single-crystalline and thin-film SrTiO$_3$.

In summary, we have investigated the low-frequency dielectric and optical properties of ferroelectric SrTiO$_3$. By use of THz-TDS, we characterized the far-IR power absorption, refractive index, and complex dielectric constant of powder form SrTiO$_3$ in the frequency range from 0.2 to 2.0 THz with the results well fit by the classical pseudo-harmonic model. The THz-TDS study implicates that the dielectric constant of SrTiO$_3$ is highly correlated with the lowest optical soft phonon mode at 90.0 cm$^{-1}$, directly measured by the Raman spectroscopy. The combination of THz-TDS and Raman light scattering is demonstrated as a powerful approach to explore the low-frequency dielectric properties associated with optical phonon resonances in semiconductors and composite materials. The comparison among different forms of SrTiO$_3$ reveals that the powder form SrTiO$_3$ exhibits similar low-frequency dielectric properties with that of single-crystalline and thin-film SrTiO$_3$.

The authors acknowledge the efforts of A. K. Azad. This work was partially supported by the National Science Foundation and the Oklahoma EPSCoR for the National Science Foundation.

**References**

1. A. S. Barker, Jr. and M. Tinkham, Phys. Rev. **125**, 1527 (1962).

2. W. G. Spitzer, R. C. Miller, D. A. Kleinman, and L. E. Howarth, Phys. Rev. **126**, 1710 (1962).

3. T. Ostapchuk, . Petzelt, V. Železný, A. Pashkin, J. Pokorný, I. Drbohlav, R. Kužel, D. Rafaja, B. P. Gorshunov, M. Dressel, Ch. Ohly, S. Hoffmann-Eifert, and R. Waser, Phys. Rev. B **66**, 235406 (2002).

4. A. A. Sirenko, C. Bernhard, A. Golnik, A. M. Clark, J. Hao, W. Si, and X. X. Xi, Nature **404**, 373 (000).

5. P. Kužel, F. Kadlec, H. Němec, R. Ott, E. Hollmann, and N. Klein, Appl. Phys. Lett. **88** 102901 (2006).

6. A. K. Tagantsev, V. O. Sherman, K. F. Astafiev, J. Venkatesh, and N. Setter, J. Electroceram **11**, 5 (2003).

7. M. Misra, K. Kotani, I. Kawayama, H. Murakami, and M. Tonouchi, Appl. Phys. Lett. **87**, 182909 (2005).

8. D. Grischkowsky, S. Keiding, M. van Exter, and Ch. Fattinger, J. Opt. Soc. Am. B **7**, 2006 (1990).

9. A. K. Azad, J. G. Han, and W. Zhang, Appl. Phys. Lett. **88**, 021103 (2006).

10. J. Han, F. Wan, Z. Zhu Y. Liao, T. Ji, M. Ge, and Z. Zhang, Appl. Phys. Lett. **87**, 172107 (2005).

11. J. Han, Z. Zhu, S. Ray, A. K. Azad, W. Zhang, M. He, S. Li, and Y. Zhao, Appl. Phys. Lett. **89**, 031107 (2006).

12. L. Thamizhmani, A. K. Azad, J. Dai, and W. Zhang, Appl. Phys. Lett. **86**, 131111 (2005).

13. T. Yamaguchi, M. Sakai, and N. Saito, Phys. Rev. B **32**, 2126 (1985).

14. M. Balkanski, *Optical Properties of Solids*, edited by F. Abelès (North-Holland, New York, 1972), Chap. 8.

15. J. L. Servoin, Y. Luspin, and F. Gervais, Phys. Rev. B **22**, 5501 (1980).

16. J. D. Axe, Phys. Rev. **157**, 429 (1967).

17. C. H. Perry, B. N. Khanna, and G. Rupprecht, Phys. Rev. **135**, A408 (1964).

18. J. Petzelt, T. Ostapchuk, I. Gregora, I. Rychetský, S. Hoffmann-Eifert, A. V. Pronin, Y. Yuzyuk, B. P. Gorshunov, S. Kamba, V. Bovtun, J. Pokorný, M. Savinov, V. Porokhonskyy, D. Rafaja, P. Vaněk, A. Almeida, M. R. Chaves, A. A. Volkov, M. Dressel, and R. Waser, Phys. Rev. B **64**, 184111 (2001).

19. A. Tkach, P. M. Vilarinho, A. L. Kholkin, A. Pashkin, S. Veljko, and J. Petzelt, Phys. Rev. B **73**, 104113 (2006).

**Figure Captions**

**FIG. 1.** (a) THz-TDS measured power absorption α (open circles) and the theoretical fitting through the classical pseudo-harmonic model and the Bruggeman effective medium approximation (solid curve); (b) measured refractive index (open circles) and the theoretical fitting (solid curve).

**FIG. 2.** Complex dielectric constant of $SrTiO_3$: (a) measured real part of dielectric constant $\varepsilon_r$ (open circles) and the theoretical fitting (solid curve); (b) measured imaginary dielectric constant $\varepsilon_i$ (open circles) and the theoretical fitting (solid curve).

**FIG. 3.** (Color online) Completed low-frequency optical response of $SrTiO_3$: the open circles are measured by THz-TDS in the frequency region 6.7 to 66.7 $cm^{-1}$, and the solid curve is the Raman spectrum recorded from 67.0 to 1000.0 $cm^{-1}$. The inset shows the detailed feature of the measured TO1 mode.

**FIG. 4.** (Color online) Comparison of THz-TDS results of (a) power absorption, and (b) refractive index of pure powder $SrTiO_3$ (squares) with that of single-crystalline (circles) and thin film (crosses, obtained from Ref. [7]) $SrTiO_3$. The solid lines represent theoretical fit.

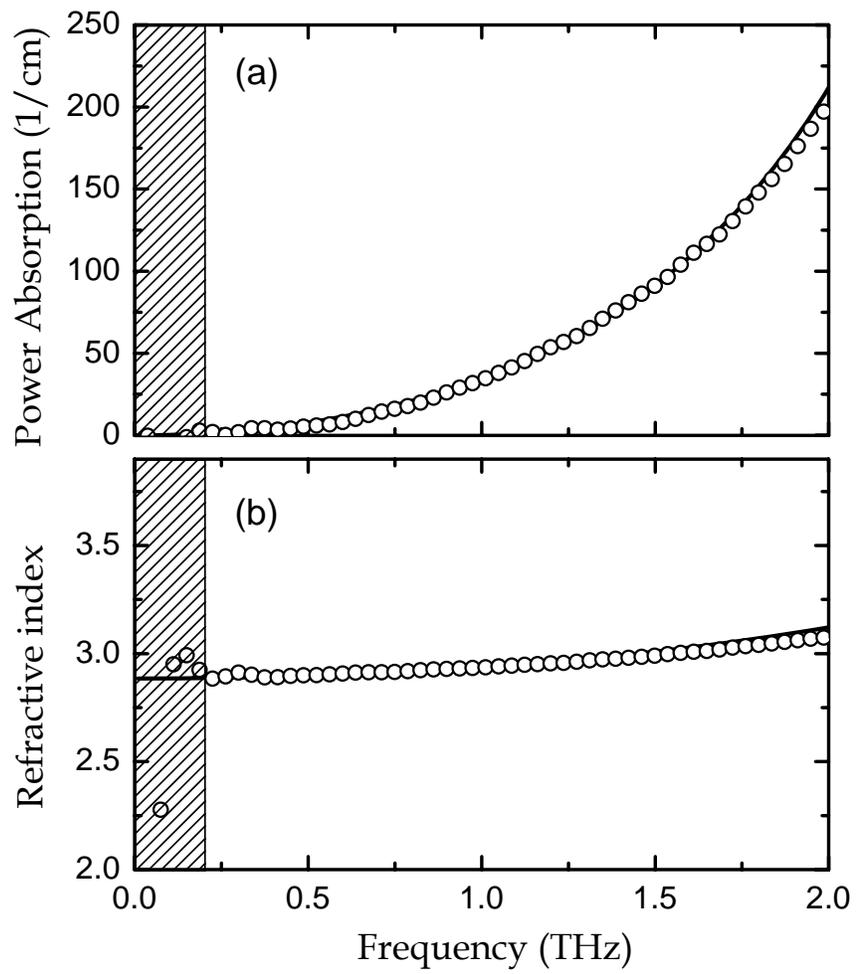

**FIG. 1.**
**Han** *et al.*

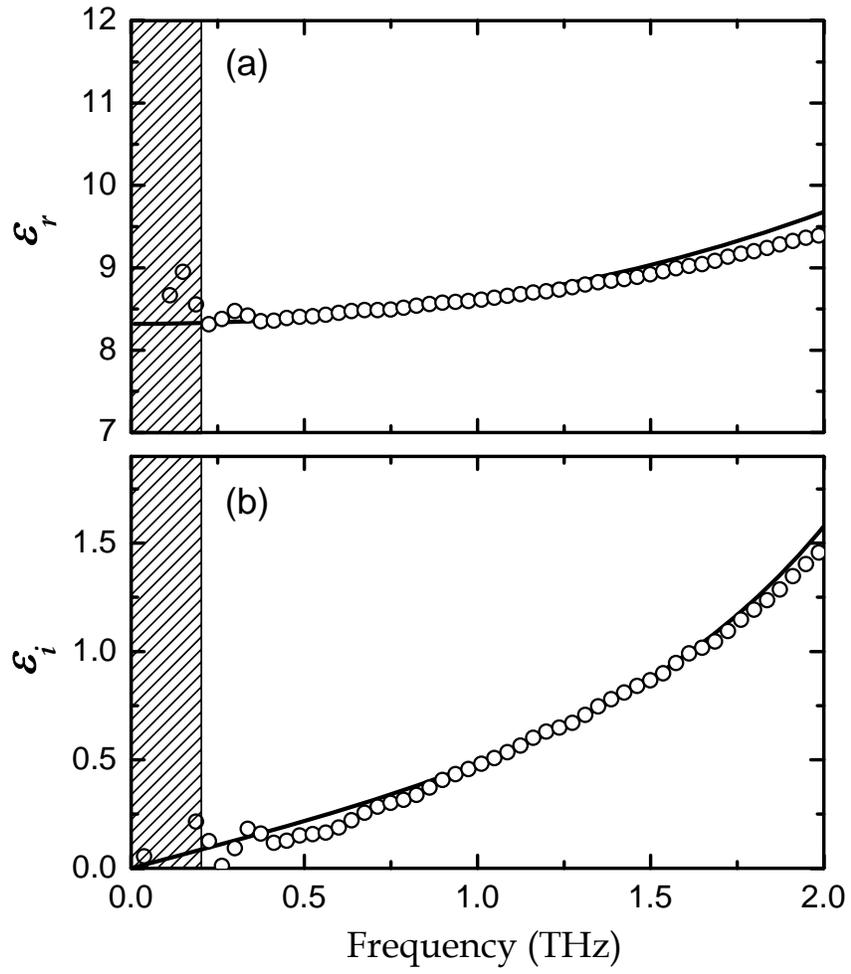

**FIG. 2.**
**Han** *et al.*

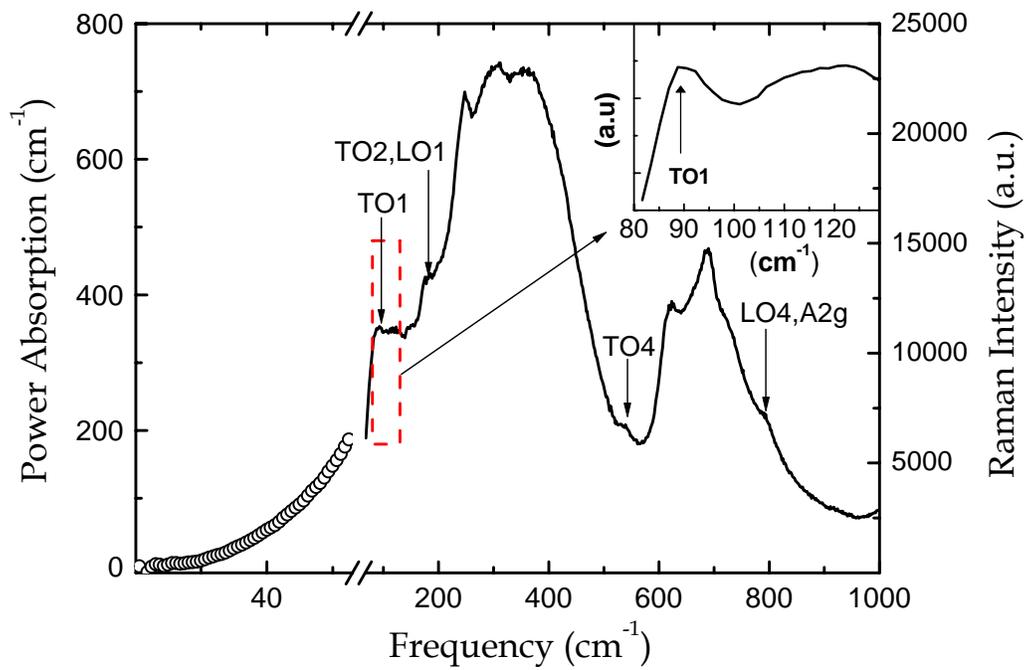

**FIG. 3.**
**Han** *et al.*

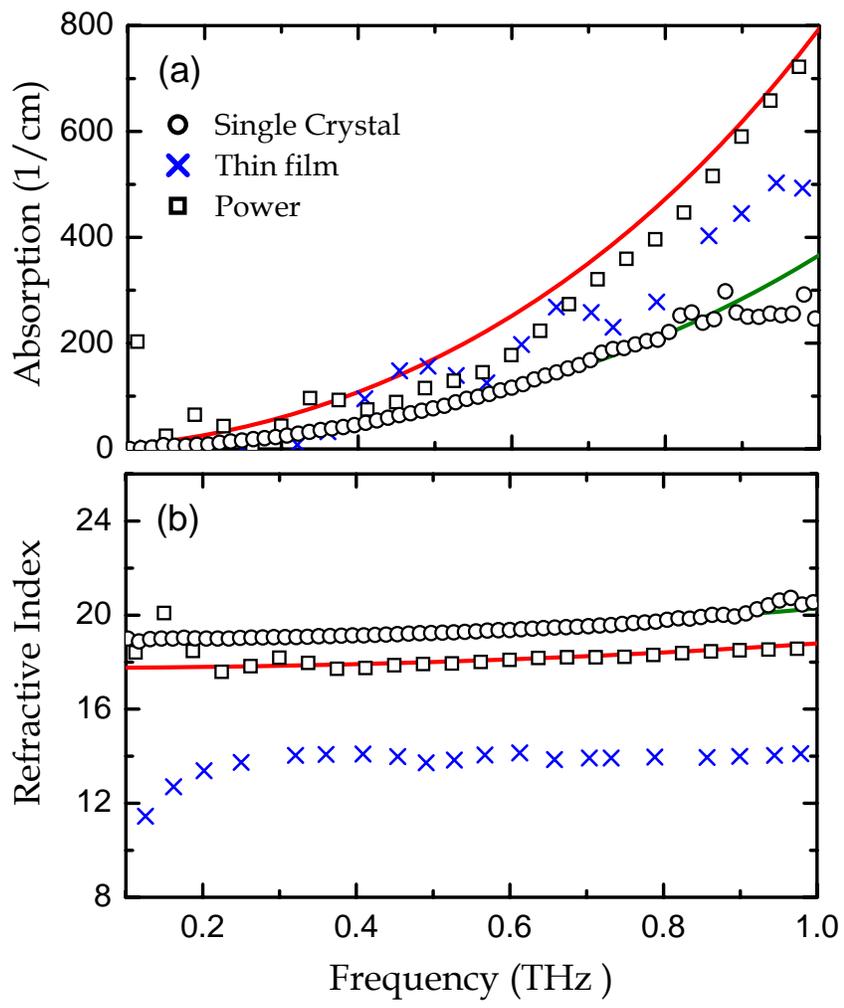

**FIG. 4.**
**Han** *et al.*

14